\documentclass[11pt]{article}
\usepackage{amsmath,amssymb}
\usepackage{graphicx}
\usepackage{slashed}
\usepackage[margin=2.25cm]{geometry}
\usepackage[blocks]{authblk}
\usepackage{caption}
\usepackage{subcaption}
\usepackage[compat=1.1.0]{tikz-feynman}
\usepackage{cite}
\usepackage{xcolor,comment}
\usepackage{ulem}

\newcommand{\ie}{\textit{i.e.,}}
\newcommand{\eg}{\textit{e.g.,}}

\date{\today}
\title{{\sf Numerically Computing  Finite Temperature Loop Integrals using pySecDec}}

\author{D.~Harnett\thanks{derek.harnett@shaw.ca}}
\author{Siyuan Li\thanks{siyuan.li@usask.ca}}
\author{T.G.~Steele\thanks{tom.steele@usask.ca}}

\affil{Department of Physics and Engineering Physics, University of Saskatchewan, Saskatoon, SK,
S7N~5E2, Canada}

\begin{document}

\maketitle

\begin{abstract}
Finite-temperature quantum  field theory provides the foundation for many important phenomena in the Standard Model and extensions, including phase transitions, baryogenesis, and gravitational waves.   
Methods are developed to enable application of  
pySecDec (a Python-language-based package designed for numerical calculation of dimensionally-regulated loop integrals) to numerically evaluate finite-temperature loop integrals in the imaginary time (Matsubara) formalism.   These  methods consist of two main elements: an inverse Wick rotation that converts a finite-temperature loop integral into a form applicable to pySecDec, and asymptotic techniques to regulate and accelerate convergence of the Matsubara frequency summations.  Numerical pySecDec evaluation of finite-temperature, two-point and three-point, one-loop topologies for scalar fields  is  used to illustrate and validate these new methodologies.  Advantages of these finite-temperature pySecDec numerical methods are illustrated by the inclusion of multiple mass and external momentum scales.  
\end{abstract}

\section{Introduction}\label{intro}
Finite-temperature quantum field theory (see \eg\ Refs.~\cite{das,kaputsa,Laine:2016hma,Ayala:2016vnt} for reviews) provides the foundation for many important phenomena in the Standard Model and beyond.  In particular, finite temperature effects in the effective potential for studying phase transitions  (see \eg\ Refs.~\cite{Aarts:2023vsf,Quiros:1999jp,Croon:2023zay,Athron:2023xlk} for reviews, applications to 
the Standard Model~\cite{Carrington:1991hz,Espinosa:1992kf} and extensions~\cite{Espinosa:2011ax,Huang:2020bbe})  are an essential ingredient  for studying baryogenesis (see \eg\ Refs.~\cite{Croon:2023zay,Morrissey:2012db} for reviews) and gravitational waves (see \eg\ Refs.~\cite{Croon:2023zay,Athron:2023xlk,Caprini:2019egz} for reviews).  

Finite-temperature quantum field theory  in the imaginary time (Matsubara) formalism amounts to a simple modification of zero-temperature propagators, replacing the temporal component of each four-momentum with a discretized Matsubara frequency~\cite{Matsubara:1955ws}.  The additional energy scale associated with temperature leads to greater calculational challenges in evaluating loop integrals, particularly in models with multiple mass scales.  It is therefore valuable to develop new methods for evaluating finite-temperature loop integrals to enable the study of increasingly elaborate extensions of the Standard Model.

The computational program pySecDec~\cite{Borowka:2017idc,Heinrich:2023til}
implements sector decomposition 
methods~\cite{Heinrich:2008si} to numerically calculate dimensionally-regularized integrals.
pySecDec  draws upon FORM~\cite{Vermaseren:2000nd,Kuipers:2013pba,Ruijl:2017dtg},
GSL~\cite{galassi}, and the CUBA library~\cite{Hahn:2004fe,Hahn:2014fua}.
It has previously  been demonstrated that  pySecDec can be adapted to 
QCD sum-rule calculations at leading- and next-to-leading order~\cite{Esau:2018gdp,Ray:2022fcl}. 
In this paper we develop  methods to enable application of pySecDec to numerically evaluate finite-temperature loop integrals.    One-loop bosonic two-point and three-point finite-temperature loop integrals (such as could emerge in scalar field theories)  are used to develop and benchmark these new methodologies.  As outlined below, two methodological elements are needed to apply pySecDec to finite-temperature loop integrals.   The first methodological element is an inverse Wick rotation needed to convert a finite-temperature
loop integral into a form applicable to pySecDec, and the second element is asymptotic techniques that accelerate or regulate the convergence of the sum over Matsubara frequencies.  

In Section~\ref{three-point-section}, three-point functions are first examined to establish the methodology in cases where the corresponding zero-temperature loop integral converges, but is overall complicated by the presence of multiple mass scales.   Then, in Section~\ref{two_point_sec}, these methods are extended to the case where dimensional regularization is needed for the corresponding zero-temperature loop integral.  A summary and discussion of the new pySecDec finite-temperature loop integration methodology  is provided in Section~\ref{conclusion_section}.

\section{The Finite-Temperature Three-Point Function}\label{three-point-section}
\begin{figure}[ht]
        \centering
        \begin{tikzpicture}
        \begin{feynman}
            \vertex (a);
            \vertex (b) [above right=2cm of a];
            \vertex (f1) [right=1.5cm of b];
            \vertex (f2) [above=2cm of f1];
            \vertex (f3) [above=2cm of f2];
            \vertex (c) [right=3cm of b];
            \vertex (d) [below right=2cm of c];
            

            \diagram* {
              (a) -- [double,double distance=0.3ex,with arrow=0.5,arrow size=0.13em, edge label=\(p_{1E}\)] 
              (b) -- [fermion, bend left, looseness=1.5, 
              edge label=\({k, m_3}\)] 
              (f2) -- [fermion, bend left, looseness=1.5, 
              edge label=\({k+p_{2E}, m_1}\)] 
              (c) -- [fermion, bend left, looseness=1.5, 
              edge label=\({k-p_{1E}, m_2}\)] 
              (b),
              (c) -- [double,double distance=0.3ex,with arrow=0.5,arrow size=0.13em, edge label=\(p_{1E}+p_{2E}\)] (d),
              (f3) -- [double,double distance=0.3ex,with arrow=0.5,arrow size=0.13em, edge label=\(p_{2E}\)] (f2),
            };
            \draw[fill=black] (b) circle(0.7mm);
            \draw[fill=black] (f2) circle(0.7mm);
            \draw[fill=black] (c) circle(0.7mm);

        \end{feynman}
    \end{tikzpicture}
    \caption{\label{fig:3pt_feynman_diagram} The 3-point function Feynman diagram 
    where the double lines represent the total incoming momenta of the external fields 
    within the model of interest  
    (\eg\ single field for a $\phi^3$ interaction or two fields for a $\phi^4$ interaction).}
    \end{figure}
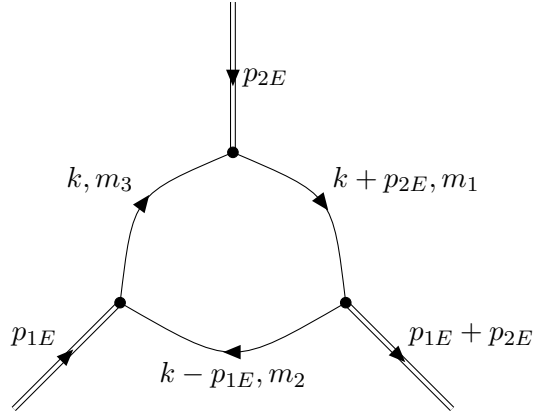

In the Matsubara formalism, the three-point vertex function for scalar fields 
in four spacetime dimensions at Euclidean external momenta 
$p_{1E}=(p^0_{1E},\,\vec{p}_1)$ and $p_{2E}=(p^0_{2E},\,\vec{p}_2)$ for
propagator masses $\{m_i\}_{i=1}^3$ and finite inverse temperature $\beta=1/T$ 
is given by (see \eg\ Ref.~\cite{das})
\begin{multline}\label{vertex_fn_beta}
    \Gamma_T\left(p_{1E},p_{2E}\right) =
    \frac{1}{\beta}\sum_{n=-\infty}^{\infty}\int\frac{d^{3} k}{(2\pi)^{3}}
    \frac{1}{\left(\omega_n^2 + |\vec{k}|^2 + m_3^2\right) \left( \left(\omega_n+p^0_{2E}\right)^2 + |\vec{k}+\vec{p}_2|^2 + m_1^2\right)}\\
    \times\frac{1}{\left( \left(\omega_n-p^0_{1E}\right)^2 + |\vec{k}-\vec{p}_{1}|^2 + m_2^2\right)}
\end{multline}
where
\begin{equation}\label{matsubara_freqs}
    \omega_n = \frac{2n\pi}{\beta}
\end{equation}
are the bosonic Matsubara frequencies (or energies)~\cite{Matsubara:1955ws}.
Note that $p^0_{1E}$ and $p^0_{2E}$ must also be Matsubara frequencies;
the subscript ``E'' has been introduced for notational consistency 
between~(\ref{vertex_fn_beta}) and its zero-temperature limit~(\ref{vertex_fn_zero_eucl}) below.
The Feynman diagram of the three-point vertex function Eq.~\eqref{vertex_fn_beta} is shown in 
Fig.~\ref{fig:3pt_feynman_diagram}, where the external momenta can be interpreted as arising 
from either a single scalar field (\eg\ a $\phi^3$ interaction) or multiple fields 
(\eg\ a $\phi^4$ interaction).
Defining the spatial integral
\begin{equation}\label{spatial_integral}
  \mathcal{S}\left(\vec{p}_1,\vec{p}_2,\Delta_1,\Delta_2,\Delta_3\right) = \int\frac{d^{3} k}{(2\pi)^{3}}
  \frac{1}{\left(|\vec{k}|^2+\Delta_3\right)\left(|\vec{k}+\vec{p}_2|^2+\Delta_1\right)\left(|\vec{k}-\vec{p}_1|^2+\Delta_2\right)}
\end{equation}
lets us write~(\ref{vertex_fn_beta}) as
\begin{equation}\label{vertex_fn_beta_S}
  \Gamma_T\left(p_{1E},p_{2E}\right) = \frac{1}{\beta}\sum_{n=-\infty}^{\infty}
  \mathcal{S}\left(\vec{p}_1, \vec{p}_2, \omega_n^2+m_3^2, \left(\omega_n+p^0_{2E}\right)^2+m_1^2, \left(\omega_n-p^0_{1E}\right)^2+m_2^2\right).
\end{equation}
Note that the spatial integral~(\ref{spatial_integral}) converges as does the
series in~(\ref{vertex_fn_beta_S}); thus, it is not necessary to use dimensional 
regularization in this (four-dimensional spacetime) case.
Convergent or not, however, pySecDec can be used to efficiently numerically evaluate
integrals like that of~(\ref{spatial_integral}). 
Section~\ref{two_point_sec} below presents an application for which dimensional regularization is necessary.

To apply pySecDec to the spatial integral~(\ref{spatial_integral}),
for an arbitrary spatial momentum $\vec{p}$, we define a corresponding Minkowski $\vec{p}_{M}$ by
\begin{gather}
  p^1_M = i p^1 \label{reverse_wick_rotation}\\
  p^i_M = p^i\ \text{for}\ i\neq 1. \label{reverse_wick_rotation_2}
\end{gather}
The transformation~(\ref{reverse_wick_rotation})--(\ref{reverse_wick_rotation_2}) 
can be thought of as an inverse Wick rotation,
but on the first spatial components rather than on the temporal components.
Then, defining $\mathcal{P}$ such that 
\begin{equation}\label{P_defn}
  \mathcal{S}\left(\vec{p}_{1},\vec{p}_{2},\Delta_1,\Delta_2,\Delta_3\right) =
    \mathcal{P}\left(\vec{p}_{1M},\vec{p}_{2M},\Delta_1,\Delta_2,\Delta_3\right)
\end{equation}
gives
\begin{multline}\label{spatial_integral_mink}
  \mathcal{P}\left(\vec{p}_{1M},\vec{p}_{2M},\Delta_1,\Delta_2,\Delta_3\right) =
  i \int\frac{d^{3} k_M}{(2\pi)^{3}}\frac{1}{\left(\vec{k}_M\cdot\vec{k}_M - \Delta_3 + i 0^{+}\right)
    \left((\vec{k}_M + \vec{p}_{2M})\cdot(\vec{k}_M + \vec{p}_{2M}) - \Delta_1 + i 0^{+}\right)}\\
    \times\frac{1}{\left((\vec{k}_M - \vec{p}_{1M})\cdot(\vec{k}_M - \vec{p}_{1M}) - \Delta_2 + i 0^{+}\right)}
\end{multline}
where we have used~(\ref{spatial_integral}) and (\ref{reverse_wick_rotation})--(\ref{P_defn}).
Throughout, we use the notation $\vec u_M\cdot\vec v_M$ to denote a Minkowski dot product of 
three-component vectors $\vec u_M,~\vec v_M$, \eg\
\begin{equation}\label{tom_mink_dot}
    \vec{k}_M\cdot\vec{k}_M = (k^1_M)^2 - (k^2)^2 - (k^3)^2.
\end{equation}%
Note that the usual $i0^+$ limit within the propagators ensures that a 
Wick rotation of~\eqref{spatial_integral_mink} in the first spatial components leads to~\eqref{P_defn}, 
and hence~Eqs.~\eqref{reverse_wick_rotation}--\eqref{reverse_wick_rotation_2}  
constitute an inverse Wick rotation.
The right-hand side of~(\ref{spatial_integral_mink}) has the form of a 
three-dimensional spacetime integral of a product of Minkowski-space propagators
and, as such, is in a form suitable for numerical evaluation using pySecDec.
Then, with
\begin{gather}
  \Delta_1 = \omega_n^2 + m_3^2\\
  \Delta_2 = \left(\omega_n + p_{2E}^0\right)^2 + m_1^2\\
  \Delta_3 = \left(\omega_n - p_{1E}^0\right)^2 + m_2^2
\end{gather}
in~(\ref{P_defn})--(\ref{spatial_integral_mink}), we can calculate 
$\Gamma_T(p_{1E}, p_{2E})$ using~(\ref{vertex_fn_beta_S}).
Thus, the inverse Wick rotation as represented in 
Eqs.~\eqref{reverse_wick_rotation}--\eqref{reverse_wick_rotation_2}
and employed in~(\ref{P_defn})--(\ref{spatial_integral_mink})
is the first methodological element needed to evaluate finite-temperature 
loop integrals with pySecDec.

The zero-temperature limit (\ie\ $\beta\to\infty$) of~(\ref{vertex_fn_beta})
is given by 
\begin{equation}\label{vertex_fn_zero_eucl}
    \Gamma_0\left(p_{1E},p_{2E}\right) = \int\frac{d^4 k_E}{(2\pi)^4}
    \frac{1}{\left(k_E^2+m_3^2\right)\left((k_E+p_{2E})^2+m_1^2\right)\left((k_E-p_{1E})^2+m_2^2\right)}
\end{equation}
where
\begin{equation}
    k_E^2 = (k^0_E)^2 + |\vec{k}|^2
    = (k^0_E)^2 + (k^1)^2 + (k^2)^2 + (k^3)^2
\end{equation}
with analogous expressions for $(k_E + p_{2E})^2$ and $(k_E - p_{1E})^2$.
Alternatively, $\Gamma_0$ can be expressed 
in terms of Minkowski momenta by defining $\Lambda_0$ through
\begin{equation}\label{Lambda_defn}
  \Gamma_0\left(p_{1E}, p_{2E}\right) = \Lambda_0\left(p_1, p_2\right)  
\end{equation}
which gives
\begin{equation}\label{vertex_fn_zero_mink}
    \Lambda_0\left(p_1, p_2\right) = i \int\frac{d^4 k}{(2\pi)^4}
    \frac{1}{\left(k^2-m_3^2 + i 0^+\right)\left((k+p_2)^2-m_1^2 + i 0^+\right)\left((k-p_1)^2-m_2^2 + i 0^+\right)}\\
\end{equation}
where 
\begin{equation}\label{eucl_to_mink}
  k^0 = i k^0_E,\ p_1^0 = i p^0_{1E},\ p_2^0 = i p^0_{2E}
\end{equation}
(note that spatial components are unaffected by such Wick rotations on temporal components)
and where
\begin{equation}
    k^2 = (k^0)^2 - |\vec{k}|^2 
    = (k^0)^2 - (k^1)^2 - (k^2)^2 - (k^3)^2
\label{tom_minkowski}
\end{equation}
with analogous expressions for $(k+p_2)^2$ and $(k-p_1)^2$.
Equations~\eqref{Lambda_defn}--\eqref{tom_minkowski} again represent an inverse Wick rotation. 
The right-hand side of~(\ref{vertex_fn_zero_mink}) is in a form suitable for numerical 
evaluation using pySecDec because the $i0^+$ limit corresponds to 
standard Minkowski-space propagators.

As an aside, we note that~(\ref{vertex_fn_zero_eucl}) can be written
in terms of the spatial integral~(\ref{spatial_integral}) as
\begin{equation}\label{vertex_fn_zero_eucl_S}
    \Gamma_0\left(p_{1E},p_{2E}\right) 
    = \int_{-\infty}^{\infty} \frac{dk^0_E}{2\pi}\,
    \mathcal{S}\left(\vec{p}_1, \vec{p}_2, (k^0_E)^2+m_3^2, 
        (k^0_E+p^0_{2E})^2+m_1^2, (k^0_E-p^0_{1E})^2+m_2^2\right).          
\end{equation}
In~(\ref{vertex_fn_zero_eucl_S}),
the integrand $\mathcal{S}$ can be numerically evaluated with pySecDec 
using~(\ref{P_defn})--(\ref{spatial_integral_mink})
for $\Delta_1 = m_1^2$, $\Delta_2 = m_2^2$, and $\Delta_3 = m_3^2$
while the outside integral over $k^0_E$ 
can be numerically evaluated using a variety of one-dimensional techniques.
Comparing $\Gamma_0$ as calculated using~(\ref{Lambda_defn})--(\ref{vertex_fn_zero_mink}),
a straightforward application of pySecDec to the entire spacetime integral,
and~(\ref{vertex_fn_zero_eucl_S}) serves as a valuable consistency check on 
the inverse Wick rotation methodology~(\ref{reverse_wick_rotation})--(\ref{spatial_integral_mink}) 
and the implementation of~(\ref{spatial_integral}). 

To identify the scaling behaviour of $\Gamma_T$ and $\Gamma_0$, 
we re-express both in terms of dimensionless quantities.
In~(\ref{vertex_fn_beta}), we introduce a dimensionless integration 
variable $\vec{\kappa}$,
\begin{equation}\label{q_defn}
    \vec{\kappa} = \frac{\beta}{2\pi}\vec{k} \implies
    \mathrm{d}^{3}\kappa = \left(\frac{\beta}{2\pi}\right)^{3} \mathrm{d}^{3}k.
\end{equation}
Similarly, we define dimensionless variables 
$\ell_1$, $\vec{q}_1$, $\ell_2$, and $\vec{q}_2$,
\begin{equation}
\label{ell_defn}
    \ell_i = \frac{\beta}{2\pi}p^0_{iE},\ 
    \vec{q}_i = \frac{\beta}{2\pi}\vec{p}_i\ \text{for}\ i\in\{1,2\},
\end{equation}
where $\ell_1$ and $\ell_2$ must be integers because $p^0_{1E}$ and $p^0_{2E}$ 
are  Matsubara frequencies. With
\begin{equation}\label{M_defn}
  M = \max\{m_i\}_{i=1}^3,
\end{equation}
we define dimensionless mass parameters $\xi_i$,
\begin{equation}\label{xi_defn}
  \xi_i = \frac{m_i}{M}.
\end{equation}
Then, we define $\tilde{\Gamma}_T$ through
\begin{equation}\label{tilde_Gamma_defn}
  \Gamma_T\left(p_{1E}, p_{2E}\right) = 
    \frac{1}{M^2} \tilde{\Gamma}_T\left(\ell_1, \vec{q}_1; \ell_2, \vec{q}_2\right).
\end{equation}
Substituting~(\ref{q_defn})--(\ref{xi_defn}) into~(\ref{vertex_fn_beta}), 
we find, using~(\ref{tilde_Gamma_defn}), 
\begin{multline}\label{vertex_fn_beta_dimless}
    \tilde{\Gamma}_T\left(\ell_1, \vec{q}_1; \ell_2, \vec{q}_2\right) =
    \frac{a^2}{2\pi} \sum_{n=-\infty}^{\infty}\int\frac{d^{3} \kappa}{(2\pi)^{3}}
    \frac{1}{\left(n^2 + |\vec{\kappa}\,|^2 + a^2\xi_3^2\right)
    \left((n+\ell_2)^2 + |\vec{\kappa}+\vec{q}_2|^2 + a^2\xi_1^2\right)}\\
    \times\frac{1}{\left((n-\ell_1)^2 + |\vec{\kappa}-\vec{q}_{1}|^2 + a^2\xi_2^2\right)}
\end{multline}
where
\begin{equation}\label{a_defn}
    a = \frac{\beta M}{2\pi}
\end{equation}
is dimensionless. Therefore, $\tilde{\Gamma}_T$ is also dimensionless.

Next for $\Gamma_0$, in~(\ref{vertex_fn_zero_eucl}), we apply the change of variables
\begin{equation}\label{q_defn_zero}
  \kappa_E = \frac{k_E}{M} \implies
  \mathrm{d}^{4}\kappa_E = \frac{1}{M^4} \mathrm{d}^{4}k_E
\end{equation}
where $M$ is defined in~(\ref{M_defn}).
Analogous to~(\ref{tilde_Gamma_defn}), we define $\tilde{\Gamma}_0$ through
\begin{equation}\label{tilde_Gamma_zero_defn}
  \Gamma_0\left(p_{1E}, p_{2E}\right) = 
    \frac{1}{M^2} \tilde{\Gamma}_0\left(\ell_1, \vec{q}_1; \ell_2, \vec{q}_2\right)
\end{equation}%
where $\ell_i$ and $\vec{q}_i$ are defined in~(\ref{ell_defn}).
Substituting~(\ref{ell_defn}), (\ref{xi_defn}), and~(\ref{q_defn_zero}) 
into~(\ref{vertex_fn_zero_eucl}), we find, using~(\ref{tilde_Gamma_zero_defn}),
\begin{multline}\label{vertex_fn_dimless}
    \tilde{\Gamma}_0\left(\ell_1, \vec{q}_1; \ell_2, \vec{q}_2\right) =
    \int\frac{d^4 \kappa_E}{(2\pi)^4}
    \frac{1}{\left[\left(\kappa^0_E\right)^2 + \left|\vec{\kappa}\right|^2 + \xi_3^2\right]
    \left[\left(\kappa^0_E+\frac{\ell_2}{a}\right)^2 + \left|\vec{\kappa} + \frac{\vec{q}_2}{a}\right|^2 + \xi_1^2\right]}
    \\ 
    \times\frac{1}{\left[\left(\kappa^0_E-\frac{\ell_1}{a}\right)^2 + \left|\vec{\kappa} - \frac{\vec{q}_1}{a}\right|^2 + \xi_2^2\right]}
\end{multline}
which shows that $\tilde{\Gamma}_0$ is dimensionless.
Strictly speaking, in~(\ref{tilde_Gamma_zero_defn})--(\ref{vertex_fn_dimless}),
$\ell_1$ and $\ell_2$ need not be integers.
Also note that \eqref{vertex_fn_dimless} can be put into a Euclidean four-dimensional  
covariant form by associating $\ell_i/a$ and $\vec q_i/a$ for each $i$ with 
the temporal and spatial components respectively of a four-dimensional Euclidean vector.

The inverse Wick rotation methodology introduced above can be applied in a straightforward way to the rescaled integrals~\eqref{tilde_Gamma_defn} and~\eqref{vertex_fn_beta_dimless}. It is also important to emphasize that  pySecDec requires input of numeric values for all parameters within the loop integrals, and hence (apart from the scaling arguments), the masses $m_i$, inverse temperature $\beta$, and external momenta 
$p_{1E}=(p^0_{1E},\,\vec{p}_1)$ and $p_{2E}=(p^0_{2E},\,\vec{p}_2)$ are needed as numeric inputs to pySecDec.

\begin{figure}[htb]
\centering
\includegraphics[scale=1]{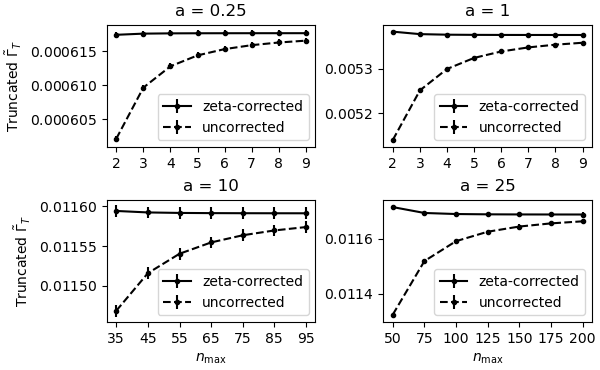}
\caption{\label{plot-convergence} The right-hand side of~(\ref{corrected}) 
    with and without the zeta-function correction versus
    $n_{\max}$ for $\ell_1=\ell_2=1$, $\vec{q}_1=\vec{q}_2=0$,
    $\xi_1 = 1$, $\xi_2 = 0.5$, and $\xi_3 = 0$ at various values of $a$.}
\end{figure}

In computing~$\tilde{\Gamma}_T$, we generally need to truncate the series
outside of $n\in\{-n_{\max},-n_{\max}+1,\ldots,n_{\max}\}$ for some $n_{\max}$.
We can, however, estimate the size of the corresponding truncation error.
Suppressing function arguments, we write
\begin{align}
    \tilde{\Gamma}_T 
    &= \sum_{n=-\infty}^{\infty}\tilde{\Gamma}_{T,n}\label{gamma_Tn}\\
    &= \sum_{n=-n_{\max}}^{n_{\max}}\tilde{\Gamma}_{T,n} +
    \sum_{n=n_{\max}+1}^{\infty}\tilde{\Gamma}_{T,n} +
    \sum_{n=-\infty}^{-(n_{\max}+1)}\tilde{\Gamma}_{T,n}
\end{align}
where, from~(\ref{vertex_fn_beta_dimless}) and~(\ref{gamma_Tn}),
\begin{multline}
    \tilde{\Gamma}_{T,n} = 
    \frac{a^2}{2\pi}\int\frac{d^3 \kappa}{(2\pi)^3}
    \frac{1}{\left(n^2 + |\vec{\kappa}\,|^2 + a^2\xi_3^2\right)
    \left((n+\ell_2)^2 + |\vec{\kappa}+\vec{q}_2|^2 + a^2\xi_1^2\right)}\\
    \times\frac{1}{\left((n-\ell_1)^2 + |\vec{\kappa}-\vec{q}_{1}|^2 + a^2\xi_2^2\right)}.
\end{multline}
If $n_{\max}$ is chosen large enough such that $|n|>n_{\max}$ implies that
$|n|\gg a \xi_i$, $|n|\gg |\ell_i|$, and $|n|\gg|\vec{q}_i|$, then
\begin{align}
    \tilde{\Gamma}_T 
    &\approx \sum_{n=-n_{\max}}^{n_{\max}}\tilde{\Gamma}_{T,n} +
    2\sum_{n=n_{\max}+1}^{\infty}
        \frac{a^2}{2\pi}\int\frac{\mathrm{d}^3 \kappa}{(2\pi)^3}
        \frac{1}{(n^2 + |\vec{\kappa}|^2)^3}\notag\\
    &= \sum_{n=-n_{\max}}^{n_{\max}}\tilde{\Gamma}_{T,n} +
        \frac{a^2}{32\pi^2}\sum_{n=n_{\max}+1}^{\infty}\frac{1}{n^3}\notag\\
    &= \sum_{n=-n_{\max}}^{n_{\max}}\tilde{\Gamma}_{T,n} +
        \frac{a^2}{32\pi^2}\left(\zeta(3) -
            \sum_{n=1}^{n_{\max}}\frac{1}{n^3}\right)\notag\\
    \implies \tilde{\Gamma}_T
    &\approx \sum_{n=-n_{\max}}^{n_{\max}}\tilde{\Gamma}_{T,n} +
        \frac{a^2}{32\pi^2}\, \zeta\left[3,n_{\max}+1 \right]
    \label{corrected}
\end{align}
where $\zeta[s,b]$ is the generalized Riemann zeta function
\begin{equation}
  \zeta\left[s,b\right]=\sum_{k=0}^\infty \frac{1}{\left(k+b\right)^s}  ~.
\end{equation}
Accordingly, we refer to the final term on the right-hand side of~(\ref{corrected})
as the zeta-function correction. In general, including zeta-function corrections
in calculations of $\tilde{\Gamma}_T$ allows for a significantly lower value of 
$n_{\max}$ to be used which speeds up computation times considerably.
This is illustrated in Figure~\ref{plot-convergence} where we plot~(\ref{corrected}) 
with and without the zeta-function correction for various values of $a$. 
The corrected versions converge more quickly than the uncorrected, 
and, hence, use of the large-$n$ asymptotic form of the finite-temperature 
loop integrals accelerates convergence of the Matsubara frequency summation. 
This asymptotic technique is the second element of our pySecDec methodology for 
finite-temperature loop integrals. In Section~\ref{two_point_sec}, 
asymptotic techniques are needed to regulate the Matsubara sum 
in addition to accelerating convergence. 
All calculations of $\tilde{\Gamma}_T$ in the rest of this section include 
the zeta-correction.

\begin{figure}[htb]
\centering
\includegraphics[scale=1]{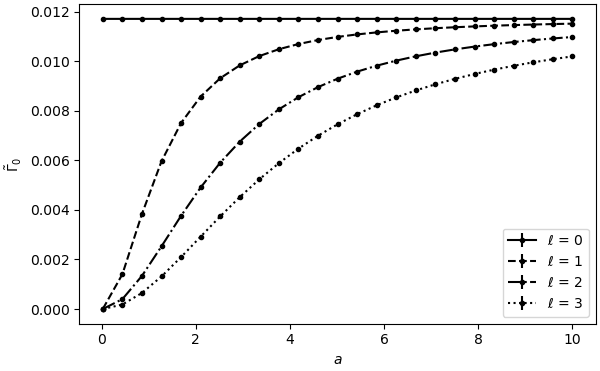}
\caption{\label{plot:Gamma_zero_temp} The dimensionless, zero-temperature 
vertex function $\tilde{\Gamma}_0$ versus $a$ 
for $\vec{q}_1 = \vec{q}_2 = 0$,
$\xi_1 = 1$, $\xi_2 = 0.5$, and $\xi_3 = 0$
at various values of $\ell=\ell_1=\ell_2$. 
Error bars  reflecting the  numerical uncertainty as determined by pySecDec are not visible.}
\end{figure}

\begin{figure}[htb]
\centering
\includegraphics[scale=1]{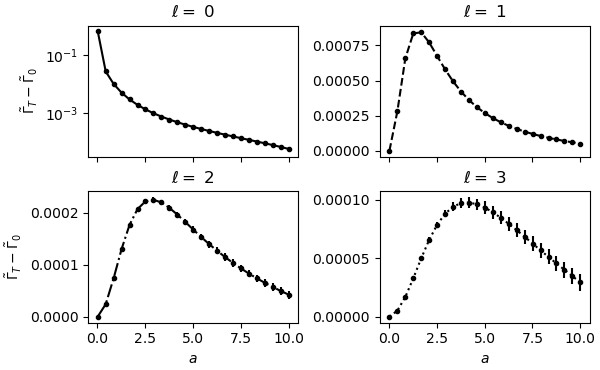}
\caption{\label{plot:Gamma_correction} The dimensionless, finite-temperature 
vertex function correction  $\tilde{\Gamma}_T - \tilde{\Gamma}_0$ 
versus $a$ for $\vec{q}_1 = \vec{q}_2 = 0$,
$\xi_1 = 1$, $\xi_2 = 0.5$, and $\xi_3 = 0$
at various values of $\ell=\ell_1=\ell_2$.
Note that the vertical-axis scale of the $\ell = 0$ plot is logarithmic
whereas the others are linear.
Error bars (where visible) reflect the  numerical uncertainty as determined by pySecDec.}
\end{figure}

\begin{figure}[htb]
\centering
\includegraphics[scale=1]{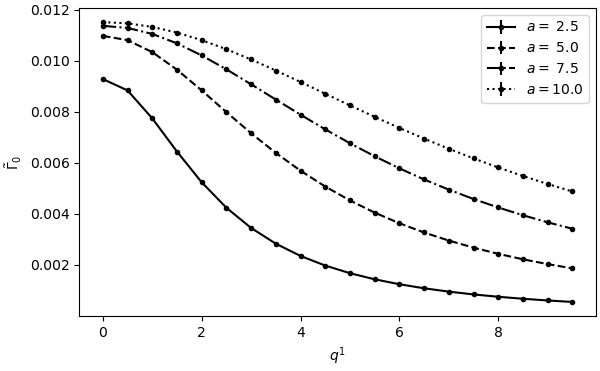}
\caption{\label{plot:Gamma_zero_temp_momentum} The dimensionless, 
zero-temperature vertex function $\tilde{\Gamma}_0$ versus $q^1$ 
for $\ell_1 = \ell_2 = 1$,
$\xi_1 = 1$, $\xi_2 = 0.5$, and $\xi_3 = 0$ at various values of $a$. 
Error bars  reflecting the  numerical uncertainty as determined by pySecDec are not visible.  }
\end{figure}

\begin{figure}[htb]
\centering
\includegraphics[scale=1]{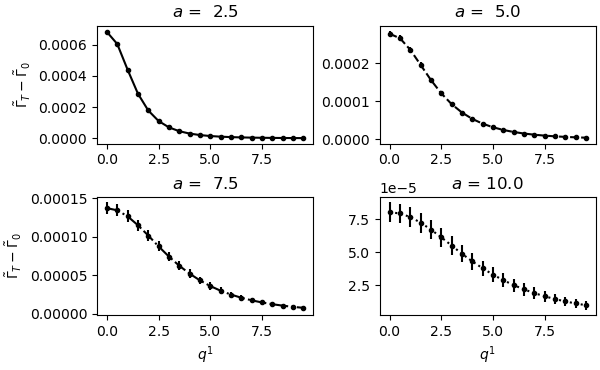}
\caption{\label{plot:Gamma_momentum_correction} The dimensionless, 
finite-temperature vertex function correction  
$\tilde{\Gamma}_T - \tilde{\Gamma}_0$ versus $q^1$ 
for $\ell_1 = \ell_2 = 1$,
$\xi_1 = 1$, $\xi_2 = 0.5$, and $\xi_3 = 0$
at various values of $a$. Error bars (where visible) reflect the  numerical uncertainty as determined by pySecDec. }
\end{figure}

To conclude this section, we show some plots of 
the dimensionless, zero-temperature vertex function $\tilde{\Gamma}_0$
(recall~(\ref{tilde_Gamma_zero_defn})--(\ref{vertex_fn_dimless}))
and the dimensionless, vertex function finite-temperature correction
$\tilde{\Gamma}_T - \tilde{\Gamma}_0$
(recall~(\ref{tilde_Gamma_defn})--(\ref{vertex_fn_beta_dimless})),
both calculated using pySecDec.
In Figure~\ref{plot:Gamma_zero_temp}, we show $\tilde{\Gamma}_0$
as a function of $a$ for several values of $\ell=\ell_1=\ell_2$.
In obtaining Figure~\ref{plot:Gamma_zero_temp}, it has been verified 
that the two approaches of~(\ref{Lambda_defn})--(\ref{vertex_fn_zero_mink}) 
and~(\ref{vertex_fn_zero_eucl_S}) are identical, 
providing a robust self-consistency check on our methodology.
In Figure~\ref{plot:Gamma_correction}, we show 
$\tilde{\Gamma}_T - \tilde{\Gamma}_0$ as a function of $a$ 
for the same values of $\ell$.
In Figure~\ref{plot:Gamma_zero_temp_momentum}, we show $\tilde{\Gamma}_0$
as a function of $q^1$ 
where $\vec{q}_1 = \vec{q}_2 = (q^1, 0, 0)$
for several values of $a$.
In Figure~\ref{plot:Gamma_momentum_correction}, we show 
$\tilde{\Gamma}_T - \tilde{\Gamma}_0$ as a function of $q^1$ 
for the same values of $a$. 
In obtaining Figures~\ref{plot-convergence}--\ref{plot:Gamma_momentum_correction} we have compared the pySecDec numerical results with an analytic calculation that is possible in the limiting case $\xi_2=\xi_3=1$ and $\vec q_1=\vec q_2=0$.  Within the remaining 
$\{a,\ell_1,\ell_2\}$-parameter space, the difference between the analytic and pySecDec numerical results are smaller than the pySecDec-provided numerical errors, providing a validation of the methodology.
The advantages and adaptability of our finite numerical pySecDec methodology are illustrated by the incredible diversity of physical scales (mass, temperature, momentum) encompassed by 
Figures~\ref{plot:Gamma_zero_temp}--\ref{plot:Gamma_momentum_correction}.

\section{The Finite-Temperature Two-Point Function}\label{two_point_sec}
In this Section, we consider a case where dimensional regularization is needed 
within the finite-temperature loop integrals, which has implications 
for the Matsubara frequency summation. 

\begin{figure}[ht]
        \centering
        \begin{tikzpicture}
        \begin{feynman}
            \vertex (a);
            \vertex (b) [right of=a];
            \vertex (c) [right=2cm of b];
            \vertex (d) [right of=c];
            \diagram* {
              (a) -- [double,double distance=0.3ex,with arrow=0.5,arrow size=0.13em, edge label=\(p\)] (b)
                  -- [fermion, half left, edge label=\({k+p, m_2}\)] (c)
               -- [fermion, half left, edge label=\({k, m_1}\)] (b),
              (c) -- [double,double distance=0.3ex,with arrow=0.5,arrow size=0.13em, edge label=\(p\)] (d),
      
                };
            \draw[fill=black] (b) circle(0.7mm);
            \draw[fill=black] (c) circle(0.7mm);
        \end{feynman}
    \end{tikzpicture}
    \caption{\label{fig:2pt_feynman_diagram} The 2-point function Feynman diagram 
    where the double lines represent the total incoming momenta of the external fields 
    within the model of interest (\eg\ a single field for a $\phi^3$ interaction 
    or two fields for a $\phi^4$ interaction).}
    \end{figure}
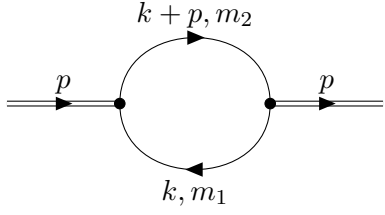

Similar to the three-point function~(\ref{vertex_fn_beta}), 
the Matsubara formalism provides an expression for the two-point function 
in four spacetime dimensions with propagator masses $\{m_i\}^2_{i=1}$ 
at Euclidean external momentum $p_E=(p^0_E,\vec{p})$
(as in Section~\ref{three-point-section}, the subscript ``E'' is used for 
notational consistency between~(\ref{2pt_fn_beta}) 
and its zero-temperature limit~(\ref{2pt_zero_temp_fcn_eucl}) below)
and finite inverse temperature $\beta=1/T$ (see \eg\ Ref.~\cite{das}):
\begin{equation}\label{2pt_fn_beta}
    \Pi_T\left(p_{E}\right) =
    \frac{1}{2\beta}\sum_{n=-\infty}^{\infty}\int\frac{d^{3} k}{(2\pi)^{3}}
    \frac{1}{\left(\omega_n^2 + |\vec{k}|^2 + m_1^2\right)\left((\omega_n+p^0_{E})^2 + |\vec{k}+\vec{p}|^2 + m_2^2\right)}
\end{equation}
where $\omega_n$, the (bosonic) Matsubara frequencies (or energies)~\cite{Matsubara:1955ws}, 
are defined in~(\ref{matsubara_freqs}) and where~(\ref{2pt_fn_beta}) 
contains a diagrammatic symmetry factor of $\frac{1}{2}$. 
Note that $p_E^0$ must be a Matsubara energy.
The Feynman diagram corresponding to~(\ref{2pt_fn_beta})
is shown in Figure~\ref{fig:2pt_feynman_diagram} where, again, the external momentum 
can be interpreted as arising from either a single scalar field (\eg\ 
a $\phi^3$ interaction) or multiple fields, \eg\ a $\phi^4$ interaction).
In anticipation of the need to use dimensional regularization in $D=4-2\epsilon$ 
spacetime dimensions, we define the spatial integral
\begin{equation}\label{2pt_spatial_integral}
    I\left(\vec{p},\Delta_1,\Delta_2\right) = \frac{\nu^{2\epsilon}}{2\beta} \int\frac{d^{D-1} k}{(2\pi)^{D-1}}
    \frac{1}{\left(|\vec{k}|^2 + \Delta_1\right)\left(|\vec{k}+\vec{p}\,|^2 + \Delta_2\right)}
\end{equation}
where $\nu$ is the renormalization scale. 
[Unlike in~(\ref{spatial_integral}), here, we absorb a factor of $\beta^{-1}$ 
into the definition of the spatial integral~(\ref{2pt_spatial_integral}).]
Using~\eqref{2pt_spatial_integral} lets us express~\eqref{2pt_fn_beta} as 
\begin{equation}\label{Pi_T_as_sum}
    \Pi_T(p_{E}) =
    \sum_{n=-\infty}^{\infty} I\left(\vec{p},\omega_n^2 +  m_1^2,(\omega_n+p^0_{E})^2 + m_2^2\right).
\end{equation}
Similarly, the dimensionally-regularized, zero-temperature ($\beta\to\infty$) 
two-point function is given by
\begin{equation}\label{2pt_zero_temp_fcn_eucl}
    \Pi_0\left(p_{E}\right) = \frac{\nu^{2\epsilon}}{2}\int\frac{d^D k_E}{(2\pi)^D} \frac{1}{\left(k_E^2+m_1^2\right)\left((k_E+p_{E})^2+m_2^2\right)}
\end{equation}
or, in terms of Minkowski momenta, by 
\begin{equation}\label{2pt_zero_temp_fcn_mink}
    \Pi_0(p) = -i\frac{\nu^{2\epsilon}}{2} \int\frac{d^D k}{(2\pi)^D} \frac{1}{\left(k^2-m_1^2+i0^+\right)\left((k+p)^2-m_2^2+i0^+\right)},
\end{equation}
where $p^0_E$ and $p^0$ as well as $k^0_E$ and $k^0$ are related as in~(\ref{eucl_to_mink}).
Throughout this section, we denote the two-point, bosonic correlator as $\Pi$ 
regardless of the particular choice of function argument(s) such as 
$p_E$ in~(\ref{2pt_zero_temp_fcn_eucl}) or $p$ in~(\ref{2pt_zero_temp_fcn_mink}).

Compared to the three-point function, calculation of the two-point function
is complicated by the fact that both~(\ref{2pt_fn_beta}) and~(\ref{2pt_zero_temp_fcn_eucl}) 
diverge in four spacetime dimensions.
To see explicitly how the divergence in~(\ref{2pt_fn_beta}) arises and can be dimensionally-regulated, 
we consider the large-$|n|$ limit of~(\ref{2pt_spatial_integral})
with $\Delta_1 = \omega_n^2 + m_1^2$ and $\Delta_2 = (\omega_n + p^0_E)^2 + m_2^2$
where $\omega_n$ is large in magnitude compared to the  masses $m_1$, $m_2$ 
and all components of the external momentum $p_E$. 
Then,
\begin{equation}\label{2pt_spatial_large_n}
    I\left(\vec{p},\omega_n^2 +  m_1^2,(\omega_n+p^0_{E})^2 + m_2^2\right) \approx 
   \frac{\nu^{2\epsilon}}{2\beta} \int\frac{d^{D-1} k}{(2\pi)^{D-1}}
    \frac{1}{\left(|\vec{k}|^2 + \omega_n^2 \right)^2}.
\end{equation}
Using standard dimensional-regularization results (see \eg\ Refs.~\cite{Laine:2016hma,pascual})
we can rewrite~\eqref{2pt_spatial_large_n} as
\begin{equation}\label{2pt_large_n_approx}
     I\left(\vec{p},\omega_n^2 +  m_1^2,(\omega_n+p^0_{E})^2 + m_2^2\right) \approx
     \frac{1}{32\pi^2}\left[1+\epsilon \left(\log{\left(\frac{\beta^2\nu^2}{4\pi^2}\right)}-\gamma_E \right) \right] \frac{1}{\left(n^2\right)^{\frac{1}{2}+\epsilon}}.
\end{equation}
Analogous to~(\ref{gamma_Tn}), we write
\begin{gather}\label{Pi_Tn_defn}
    \Pi_T = \sum_{n=-\infty}^{\infty} \Pi_{T, n}\,,
    \\
   \Pi_{T, n}=I\left(\vec{p},\omega_n^2 +  m_1^2,(\omega_n+p^0_{E})^2 + m_2^2\right)\,,
\end{gather}
where, from~(\ref{Pi_T_as_sum}) and~(\ref{2pt_large_n_approx}), we have for large $|n|$ 
\begin{equation}\label{Pi_T_large_n} 
   \Pi_{T, n} \approx \frac{1}{32\pi^2}
   \left[1+\epsilon \left(\log{\left(\frac{\beta^2\nu^2}{4\pi}\right)}-\gamma_E \right) \right]\frac{1}{|n|^{1+2\epsilon}}.
\end{equation}
In the $\epsilon\to 0$ limit, we see from~(\ref{Pi_Tn_defn}) and~(\ref{Pi_T_large_n}) 
that $\Pi_T$ is divergent. 
However, dimensional regularization does  parameterize  this divergence via the zeta function. As for the three-point function, this  regularization is achieved by truncating the exact series \eqref{Pi_Tn_defn} and replacing it with the large $n$ form \eqref{Pi_T_large_n} 
for $\vert n\vert>n_{\max}$
\begin{equation}
\begin{aligned}
    \Pi_T \approx& \sum_{n=-(n_{\max}-1)}^{n_{\max}-1} \Pi_{T, n}
    +
    \frac{1}{16\pi^2}
   \left[1+\epsilon \left(\log{\left(\frac{\beta^2\nu^2}{4\pi}\right)}-\gamma_E \right) \right]\sum_{n=n_{\max}}^\infty \frac{1}{n^{1+2\epsilon}}
\\
 =&\sum_{n=-(n_{\max}-1)}^{n_{\max}-1} \Pi_{T, n}+\frac{1}{16\pi^2}
   \left[1+\epsilon \left(\log{\left(\frac{\beta^2\nu^2}{4\pi}\right)}-\gamma_E \right) \right]\zeta\left[1+2\epsilon,n_{\max} \right]
    \,.
\end{aligned}
\label{zeta_regulated_Pi_T}
\end{equation}
Using the result
\begin{equation}
    \zeta\left[1+2\epsilon,n_{\max} \right]=\frac{1}{2\epsilon}-\psi\left[n_{\max} \right]+{\cal O}(\epsilon )\,,
\end{equation}
where $\psi(z)$ is the digamma function, leads to a dimensionally-regularized 
(divergent) expression for $\Pi_T$
\begin{equation}
    \Pi_T \approx \sum_{n=-(n_{\max}-1)}^{n_{\max}-1} \Pi_{T, n}-
    \frac{1}{16\pi^2}\psi\left[n_{\max} \right]
    +\frac{1}{32\pi^2}\left[\frac{1}{\epsilon}-\gamma_E+
   \log{\left(\frac{\beta^2\nu^2}{4\pi}\right)}
    \right]    
    \,,
    \label{divergent_Pi_T}
\end{equation}
where irrelevant terms of ${\cal O}(\epsilon)$ are omitted and the approximation can be improved by increasing $n_{\max}$ so that~\eqref{2pt_large_n_approx} becomes more accurate.

Although $\Pi_T$ and $\Pi_0$ separately diverge, their difference, 
\ie\ the finite-temperature correction $\Pi_s$,
\begin{equation}\label{Pi_s}
  \Pi_s = \Pi_T - \Pi_0
\end{equation}
is finite. This behaviour is expected in mass-independent regularization schemes like dimensional regularization.
As shown in~\eqref{divergent_Pi_T}, the divergence in $\Pi_T$ comes from a series in $n$ whereas 
the divergence in $\Pi_0$ comes from an integral over $k_E$ (recall~(\ref{2pt_zero_temp_fcn_eucl})). 
We can show analytically that these two divergences cancel 
in $\Pi_s$ defined by~(\ref{Pi_s}).  
Expressing \eqref{2pt_zero_temp_fcn_eucl} in terms of the dimensionless integration variable 
$\kappa_E=k_E/M$  gives 
\begin{gather}
    \Pi_0\left(p_{E}\right) = \left(\frac{\nu^2}{M^2}\right)^\epsilon  \frac{1}{2}\int\frac{d^D \kappa_E}{(2\pi)^D} \frac{1}{\left(\kappa_E^2+\xi_1^2\right)\left((\kappa_E+q_{E})^2+\xi_2^2\right)}\,,
    \\
    \label{2pt_zero_temp_fcn_eucl_rescaled}
 M = \max\{m_i\}_{i=1}^2\,,~  \xi_i=\frac{m_i}{M}\,,~ q_E=\frac{p_E}{M}\,,~q_E^0=\frac{\ell}{a}
\end{gather}
where $a$ is defined in~\eqref{a_defn} and $\ell$ is the integer corresponding to the 
Matsubara energy $p_E^0=2\pi\ell/\beta$.  
Standard dimensional regularization methods (see \eg\ Ref.~\cite{pascual}) result in a divergence that is independent of $p_E$ and the parameters $\xi_i$
\begin{equation}
    \Pi_0\left(p_{E}\right)=\frac{1}{32\pi^2} \left[\frac{1}{\epsilon}-\gamma_E+\log{(4\pi)}
    -\log{\left(\frac{M^2}{\nu^2}\right) } 
    \right]+ f^{\rm finite}_0\left(q_{E},\xi_i\right)
    \label{Pi_0_divergent}
\end{equation}
where $f^{\rm finite}_0$ contains the remaining finite parts of $\Pi_0$.
Having analytically extracted the divergences from both $\Pi_T$ and $\Pi_0$, we
have shown explicitly that they are equal and, as expected, temperature-independent.
Thus the divergences cancel from the finite-temperature correction~\eqref{Pi_s} leading to the the result 
\begin{equation}
  \Pi_s = \sum_{n=-(n_{\max}-1)}^{n_{\max}-1} \Pi_{T, n}-
    \frac{1}{16\pi^2}\psi\left[n_{\max} \right]
    +\frac{1}{16\pi^2}\log{\left( \frac{a}{2} \right)}-f^{\rm finite}_0\left(q_{E},\xi_i\right)\,.
    \label{Pi_s_finite}
\end{equation}
Because the divergent parts cancel in $\Pi_s$, the renormalization scale also cancels 
in~\eqref{Pi_s_finite} along with the usual dimensional-regularization terms
$\gamma_E$ and $\log{(4\pi)}$.
Thus, to apply pySecDec to~\eqref{Pi_s_finite}, we use $\Pi_{T,n}$ as calculated in pySecDec, 
and extract only the finite part of $\Pi_0$ from the pySecDec calculation to give a slightly 
modified version of~\eqref{Pi_s_finite} suited to pySecDec,
\begin{equation}
  \Pi_s=  \sum_{n=-(n_{\max}-1)}^{n_{\max}-1} \Pi_{T, n}-
    \frac{1}{16\pi^2}\psi\left[n_{\max} \right]
    +\frac{1}{16\pi^2}\log{\left( \frac{a}{2} \right)}
    +\frac{1}{32\pi^2}\left[-\gamma_E+\log{(4\pi)} \right]
    -\tilde \Pi^{\rm finite}_0\left(q_{E},\xi_i\right)\,,
    \label{Pi_s_finite_pySecDec}
\end{equation}
where $\tilde\Pi_0^{\rm finite}$ is the finite part of the dimensionless 
integral~\eqref{2pt_zero_temp_fcn_eucl_rescaled} omitting the $\nu^2/M^2$ renormalization-scale pre-factor
\begin{equation}
    \tilde\Pi_0\left(q_{E},\xi_i\right) =  \frac{1}{2}\int\frac{d^D \kappa_E}{(2\pi)^D} \frac{1}{\left(\kappa_E^2+\xi_1^2\right)\left((k_E+q_{E})^2+\xi_2^2\right)}\,.
\label{tilde_Pi_0_tom}
\end{equation}
Note that the $\gamma_E$ and $\log{(4\pi)}$ terms have been restored in~\eqref{Pi_s_finite_pySecDec} 
because they cannot easily be separated out from the pySecDec finite part. 
As an important numerical benchmark, the divergent part of $\Pi_0$ should be calculated in pySecDec 
to verify the $1/(32\pi^2 \epsilon)$-dependence in~\eqref{Pi_0_divergent}.  
The only remaining task is to develop a version of $\Pi_{T,n}$ suited to pySecDec. 

As for the three-point function, we switch to dimensionless parameters.
Based on our analytic result, we can anticipate the finiteness 
of~\eqref{2pt_spatial_integral} within dimensional regularization in $D=4-2\epsilon$ spacetime dimensions.
Also, in $\Pi_{T,n}$, we can set $\epsilon=0$, introduce dimensionless 
$\vec{\kappa}$ as in~(\ref{q_defn}),
and define dimensionless quantities $\ell$, $\vec{q}$, and $a$ as follows:
\begin{equation}\label{ell_q_a_defn}
  \ell = \frac{\beta}{2\pi} p^0_E,\ 
  \vec{q} = \frac{\beta}{2\pi} \vec{p}\,, 
\end{equation}
where $a$ is defined in~\eqref{a_defn} and $\ell$ is an integer because 
$p_E^0$ is a Matsubara energy. Then,
\begin{equation}\label{2pt_pi_n}
    \Pi_{T,n}\left(\{l,\vec{q}\,\}\right) = 
    \frac{1}{4\pi} \int \frac{d^3 \kappa}{(2\pi)^3} 
    \frac{1}{\left(n^2 + |\vec{\kappa}|^2 + \xi_1 a^2\right) 
             \left( \left(n + \ell\right)^2 + |\vec{\kappa} + \vec{q}\,|^2 + \xi_2 a^2\right)}.
\end{equation}
For $\Pi_{T,n}$, we implement an inverse Wick rotation much as we did
when calculating the finite-temperature three-point function.
Following~(\ref{reverse_wick_rotation}), we define $\kappa^1_M$ and $q^1_M$ as
\begin{equation}
  \kappa^1 = -i \kappa^1_M,\ q^1 = -i q^1_M.    
\end{equation}
Then,
\begin{equation}\label{2pt_pi_n_mink}
    \Pi_{T,n}\left(\{l,\vec{q}_M\,\}\right) = 
    \frac{-i}{4\pi} \int \frac{d^3 \kappa_M}{(2\pi)^3} 
    \frac{1}{\left(\vec{\kappa}_M \cdot \vec{\kappa}_M  - \Delta_1 + i0^+\right) 
             \left( (\vec{q}_M+\vec{\kappa}_M) \cdot (\vec{q}_M+\vec{\kappa}_M) - \Delta_2 + i0^+\right)}
\end{equation}
where
\begin{gather}
    \Delta_1 = n^2 + \xi_1 a^2\\
    \Delta_2 = (n + \ell)^2 + \xi_2 a^2
\end{gather}
and, as in~(\ref{spatial_integral_mink}), dot products 
in~(\ref{2pt_pi_n_mink}) are Minkowski [see Eq.~\eqref{tom_mink_dot}]. 
As discussed in Section~\ref{three-point-section}, the right-hand side of~(\ref{2pt_pi_n_mink}) is in
a form suitable for numerical evaluation with pySecDec.
As for $\tilde\Pi_0$ in~\eqref{tilde_Pi_0_tom}, its equivalent integral in terms of Minkowski-space propagators  
(in $D=4-2\epsilon$ dimensions) is 
\begin{equation}\label{2pt_zero_dimless_mink_duplicate}
    \tilde\Pi_0\left(q,\xi_i\right) = \frac{1}{2}
    \int\frac{d^D \kappa}{(2\pi)^D} \frac{1}{\left(\kappa^2 - \xi_1a^2+i0^+\right)
    \left((\kappa + q)^2 - \xi_2a^2 +i0^+\right)}
\end{equation}
where 
\begin{equation}
    \kappa^0 = i \kappa^0_E\ \text{and}\ q^0 = i \ell. 
\end{equation}
The right-hand side of~(\ref{2pt_zero_dimless_mink_duplicate}) is also in a form suitable 
for numerical calculation using pySecDec.

Before proceeding with some numerical results for the finite-temperature correction two-point function $\Pi_s$, we summarize some key aspects associated with the pySecDec evaluation of~\eqref{Pi_s_finite_pySecDec}.  
First, analytic methods have been used to regulate the divergences in the finite-temperature Matsubara 
sum~\eqref{Pi_Tn_defn}, and it has been demonstrated that these divergences  cancel against   
the divergent zero-temperature result. As for the three-point function, these analytic methods also improve the numerical convergence of the Matsubara sum.
Second, it should be verified that the pySecDec-computed values of~\eqref{2pt_pi_n_mink} 
appearing in the sum are finite, and that the  divergent part of~\eqref{2pt_zero_dimless_mink_duplicate} as computed with pySecDec 
is $1/(32\pi^2\epsilon)$ as needed to cancel the divergences. Finally, we also note that  
in~\eqref{Pi_s_finite_pySecDec}, there is a natural cancellation of the $\gamma_E$ and $\log{(4\pi)}$  
as shown analytically in \eqref{Pi_s_finite}.

\begin{figure}[htb]
\centering
\includegraphics[scale=1]{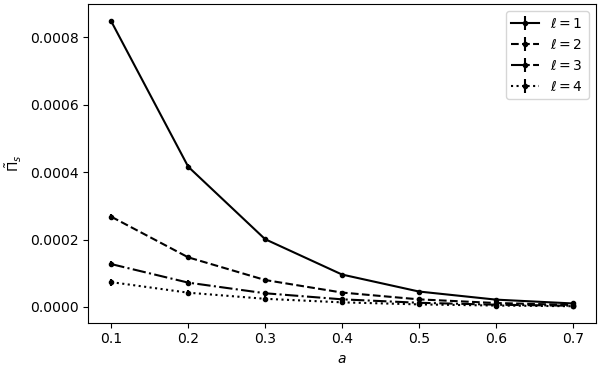}
\caption{\label{plot:Pi_correction_vs_a} The dimensionless, 
finite-temperature 2-point function correction  
${\Pi}_s$  versus $a$ 
for $\xi_1=\xi_2=1$ and $\vec{q} = 0$ 
at various values of $\ell$. Error bars (where visible) reflect the numerical uncertainty as determined by pySecDec.
} 
\end{figure}

\begin{figure}[htb]
\centering
\includegraphics[scale=1]{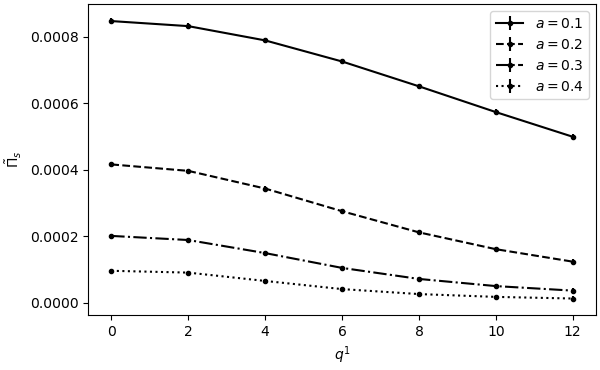}
\caption{\label{plot:Pi_correction_vs_q1} The dimensionless, 
finite-temperature 2-point function correction  
${\Pi}_s$
versus $q^1$ where $\vec{q} = (q^1,0,0)$
for $\xi_1=\xi_2=1$ and $\ell = 1$ 
at various values of $a$. 
Error bars (where visible) reflect the numerical uncertainty as determined by pySecDec. 
}
\end{figure}

As an example of results obtained from the procedure described above for numerically calculating $\Pi_s$,
we plot $\Pi_s$ as a function of $a$ at $\vec{q} = 0$ 
for several values of $\ell$ in Figure~\ref{plot:Pi_correction_vs_a}.
Also, in Figure~\ref{plot:Pi_correction_vs_q1}, we plot $\Pi_s$ 
as a function of $q^1$ where $\vec{q}=(q^1, 0, 0)$ 
at $\ell = 1$ for several values of $a$.
In obtaining these plots, it has been verified that the  case 
where $\ell=0$ and $\vec{q}=0$
agrees with the  results of Ref.~\cite{das} given in a considerably different form and approach, 
providing a robust validation of the regularization method used for the Matsubara sum.
In obtaining Figure~\ref{plot:Pi_correction_vs_a}  we have also compared the pySecDec numerical results with an analytic version 
of our calculation that is possible in the limiting case $\xi_1=\xi_2=1$ and $\vec q= 0$.  
Within the remaining $\{a,\ell\}$ parameter space, the difference between the analytic and pySecDec numerical results are smaller than the pySecDec-provided numerical errors, providing  a validation of the methodology.
As for the three-point function, the advantages and adaptability of our finite temperature numerical pySecDec methodology are illustrated by the incredible diversity of physical scales (mass, temperature, momentum) encompassed by Figures~\ref{plot:Pi_correction_vs_a}--\ref{plot:Pi_correction_vs_q1}.

\section{Discussion}\label{conclusion_section}
In this paper we have developed methods to enable application of  
pySecDec~\cite{Borowka:2017idc,Heinrich:2023til} to numerically evaluate dimensionally-regulated, finite-temperature bosonic loop integrals in the imaginary time (Matsubara) formalism. 
The methods are developed at one-loop order in four spacetime dimensions and consist of two main elements.  
The first element is an inverse Wick rotation of a spatial component
that maps a finite-temperature spatial integral into a form that enables the use of pySecDec.  
This inverse Wick rotation is a generic methodology that can easily be extended to higher-loop integrals, 
various loop topologies, and to different choices of spacetime dimension. 
The second methodological element develops asymptotic forms that can be used to regulate 
and accelerate convergence of the Matsubara sum.  
In principle, this asymptotic methodology could be extended to higher-loop calculations, 
but, of course, the analysis would become increasingly complicated 
with multiple Matsubara sums.  
Both of the methodological elements are easily adaptable to fermionic finite-temperature loop integrals.  

Two examples were used to develop and illustrate our finite-temperature pySecDec methodology.  
The finite-temperature bosonic three-point function was first considered because in four spacetime dimensions, 
the loop integrals and Matsubara sum converge.  
However, although dimensional regularization is not strictly needed, 
we find that pySecDec still provides demonstrable advantages in computational 
efficiency compared with direct numerical integration 
using standard Python functions such as {\sf tplquad }from {\sf scipy.integrate}
or Mathematica's {\sf NIntegrate}. 
We thus conclude that pySecDec is particularly well optimized to loop integrals, 
and is capable of handling the full diversity of scales that could occur in a finite-temperature system.
Despite this numerical efficiency, we also demonstrated that convergence of the Matsubara sum can be 
accelerated by using analytic methods for the asymptotic form of the finite-temperature series. 

The second example of the finite-temperature bosonic two-point function in four spacetime dimensions
has an additional complication of divergences in the Matsubara sum that we regulate using   dimensional regularization combined with  the asymptotic form of the finite-temperature series.  
We demonstrate that this  divergence is identical to that of the zero-temperature two-point function, 
and hence the finite-temperature correction~\eqref{Pi_s} is finite as expected.  An algorithm was presented for applying pySecDec to the two-point finite-temperature correction by combining analytic methods and extracting specific  portions of the zero-temperature loop integral in pySecDec. In the analysis of the two-point function, the numerical pySecDec results for the finite-temperature correction were compared against known results \cite{das} in specific  limiting cases. 

In the various figures (see Figures~\ref{plot-convergence}--\ref{plot:Gamma_momentum_correction} and Figures~\ref{plot:Pi_correction_vs_a}--\ref{plot:Pi_correction_vs_q1}) used to demonstrate our pySecDec finite-temperature loop-integral methodology, we have shown the  numerical uncertainties reported by pySecDec.  
In obtaining these figures, we have  tested the pySecDec results against limiting cases where we can perform analytic calculations,  and find  the difference between the analytic
and pySecDec numerical results are smaller than the pySecDec-provided numerical
errors, and see no evidence of numerical noise in the data generated by pySecDec.

\section*{Acknowledgments}
TGS and DH are grateful for research funding from the Natural Sciences \& Engineering Research Council of Canada (NSERC).

\end{document}